# Evanescently coupled multimode spiral spectrometer

BRANDON REDDING[†], SENG FATT LIEW[†], YARON BROMBERG, RAKTIM SARMA, HUI CAO[*]

*Department of Applied Physics, Yale University, New Haven, CT 06520*

*Corresponding author: hui.cao@yale.edu*

[†]These authors contributed equally to this work.

**We designed a high-resolution compact spectrometer based on an evanescently-coupled multimode spiral waveguide. Interference between the modes in the waveguide forms a wavelength-dependent speckle pattern which is used as a fingerprint to identify the input wavelength after calibration. Evanescent coupling between neighboring arms of the spiral results in a non-resonant broad-band enhancement of the spectral resolution. Experimentally, we demonstrated a resolution of 0.01 nm at a wavelength of 1520 nm using a 250 μm radius spiral structure. Spectra containing 40 independent spectral channels are recovered simultaneously and the operation bandwidth is significantly increased by applying compressive sensing to sparse spectra reconstruction.**

http://

## 1. INTRODUCTION

In recent years there have been rapid advances in miniaturizing spectrometers for low-cost, portable sensing applications. However, the miniaturization is often at the cost of resolving power. For example, chip-scale spectrometers based on diffractive gratings—following the same general design as bulk optics spectrometers—provide only modest spectral resolution (~0.1 nm) despite a relatively large footprint (~cm) [1,2]. As a result, a number of alternative designs for on-chip spectrometers have been developed, e.g., based on waveguide arrays [3–6], digital planar holography [7–9], dispersive photonic crystals [10,11], and micro or nano resonators [12–18]. All these designs rely on the conventional one-to-one spectral-to-spatial mapping, which is necessary for a wavelength demultiplexer but not a spectrometer. Complex spectral to spatial mapping has been explored with disordered photonic crystals, thin scattering media, random polychromat, etc for spectrometer application [19–28]. The disorder-induced scattering of light produces wavelength-dependent speckle patterns which can be used as fingerprints to identify unknown spectra. We recently utilized multiple scattering of light in a random photonic chip to obtain sub-nanometer (0.75 nm) resolution (at a wavelength of 1500 nm) with a very small (25 μm radius) footprint [22]. In contrast to single scattering or diffraction from a thin disordered material that gives a linear scaling of spectral resolution with dimension *L*, the multiple scattering in a lossless diffusive medium makes the scaling quadratic, thus the resolution increases more rapidly with *L*. However, the out-of-plane scattering loss prohibits further improvement of resolution by increasing the size of a disordered chip. Nonetheless, this general approach of using wavelength dependent speckle patterns to build a spectrometer enabled both low-loss and high-resolution using a multimode fiber in which the speckle patterns are formed by modal interference [29–32]. In the fiber spectrometer, the spectral resolution increases with the fiber length, enabling 1 pm resolution in a 100 m long fiber [32].

A straightforward on-chip implementation of the fiber spectrometer could be achieved by fabricating a long multimode waveguide. For a realistic chip size, the resolution of a straight waveguide is rather limited; however, the waveguide may be coiled in a spiral geometry to provide a long length in a small footprint [33–40]. Such spiral waveguides have been used for sensing [33,34,36], as delay lines [41], as optical comb filters [35], for supercontinuum generation [42] and for frequency stabilization [43]. Evanescent coupling between neighboring arms of the spiral was introduced to create slow-light resonances, for applications as a coupled resonant optical waveguide (CROW) [37] and an optical gyroscope [38].

However, most spiral based devices operate in a single waveguide mode which does not generate wavelength-sensitive speckle patterns, and thus could not function as a spectrometer. A Fourier-transform spectrometer based on a single-mode waveguide coiled in a spiral geometry has been demonstrated recently, but it requires an array of spiral waveguides of varying lengths, which increases the footprint drastically and does not utilize evanescent coupling between adjacent waveguides in each spiral [44].

Here, we demonstrate a chip-scale high-resolution spectrometer based on a highly multimode waveguide coiled in a spiral geometry. Interference between the waveguide modes forms a wavelength-dependent speckle pattern, which is used as a fingerprint to identify the input wavelength. The spectral resolution is greatly enhanced by introducing evanescent coupling between neighboring waveguide arms, and the enhancement is non-resonant and thus broad-band. Experimentally, we show that a spiral spectrometer with 250 μm radius can resolve two spectral lines separated by 0.01 nm in wavelength at 1520 nm. To increase the spectral bandwidth, we adopt the compressive sensing algorithm to reconstruct sparse spectra. Thanks to the complex spectral-to-spatial mapping, each spectral channel is projected to all spatial channels, making it possible to recover many frequencies with a small number of detectors.

## 2. EVANESCENTLY COUPLED SPIRAL WAVEGUIDE

The resolution of a spectrometer is dictated by the temporal spread that light experiences when propagating through the dispersive medium—in our case the multimode spiral waveguide. It is important to note that it is the temporal spread, or equivalently, the distribution of the optical path-lengths, that sets the spectrometer resolution. A long single-mode waveguide would not provide any spectral diversity for the output intensity, since all of the light travels the same distance and merely acquires a phase delay. In a multimode waveguide (as in a multimode fiber), the interference of optical paths with different length or phase delay produces an output intensity pattern that is wavelength sensitive. The sensitivity is proportional to the spread of the path-lengths, which is characterized by the difference between the shortest and longest optical paths through the waveguide. This difference is equal to the product of the physical length of the waveguide and the difference in propagation constants between the fundamental mode and the highest-order mode [30]. While increasing the length of a multimode fiber enabled us to achieve ultrafine resolution, the physical length of a multimode waveguide on-chip is limited, even when coiled tightly in a spiral geometry.

By introducing evanescent coupling between neighboring waveguides in the spiral, we could dramatically increase the spread of the optical path-lengths. For example, in an Archimedean-shaped spiral [left schematic in Fig. 1(a)], light is launched into the outer arm and propagates inwards to the center. If adjacent arms are in sufficient proximity, light is evanescently coupled between arms. When light is coupled to the inner arm, the path to the center of the spiral, where the output port is located, is shortened. Alternatively, light may be coupled to the outer arm, which lengthens the path. Hence, the actual path-length can be much shorter or longer than the physical length of the waveguide. In other words, the evanescent coupling enables light to "jump" forward or backward along the path and significantly increases the temporal spread of light.

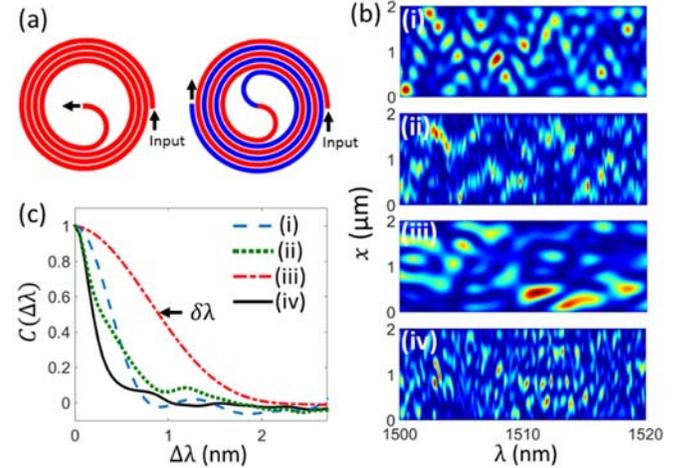

Fig. 1. Numerical modeling of the evanescently coupled multimode spiral spectrometer. (a) Schematic of a single Archimedean spiral (left) and two interleaved Archimedean spirals connected at the center by an "S-bend" waveguide (right). (b) Spectral-spatial transfer matrix for the interleaved spiral structure without evanescent coupling and mode mixing (i), with only evanescent coupling (ii), with only mode mixing (iii), with both mode mixing and evanescent coupling (iv). (c) Spectral correlation function $C(\Delta\lambda)$ for the 4 cases in (b). The spectral correlation width (HWHM of $C(\Delta\lambda)$) $\delta\lambda$ is 0.4 nm for (i), 0.3 nm for (ii), 0.9 nm for (iii), and 0.17 nm for (iv).

To further enhance the temporal spread, we considered two interleaved Archimedean spirals, which are connected at the center by an "S-bend" waveguide [right schematic in Fig. 1(a)]. Light from the input port propagates inwards through one spiral (red colored), and then outwards through the other spiral (blue colored) to the output port. Note that light in adjacent arms propagates in the opposite direction. When light is coupled to an adjacent arm, it will propagate towards the input port. A subsequent coupling will result in propagation toward the output port again. Such back-and-forth coupling will greatly shorten or lengthen the optical path length, thus increasing the temporal spread of light. Consequently, the spectral resolution will be drastically enhanced.

To verify that evanescent coupling enhances the spectral resolution, we performed numerical simulations of the interleaved Archimedean spirals. The local radius of curvature, $R$, is defined as a function of the azimuthal angle $\theta$ as $R(\theta) = R_0 - a\theta/\pi$, where $R_0$ denotes the radius of the spiral, $a$ is the center-to-center arm spacing (the sum of the waveguide width and the coupling gap) [33–35,39,40], and $\theta$ increases from $0$ to $\theta_0$. The number of spiral arms is given by $\theta_0$. The evanescent coupling is introduced to modes in neighboring waveguides that have the same order and the same propagation direction. The higher-order modes experience stronger evanescent coupling than the lower-order mode. The coupling rate is calculated by the finite-difference frequency-domain method (COMSOL) (see details in Supplement 1).

To model light propagation in the spiral waveguide, we adopted the method in refs. [36–38] (see details in Supplement 1). The waveguide is 2 µm wide and supports six guided modes in the wavelength range of 1500nm - 1520 nm. The refractive index inside the waveguide is equal to 2.86, and outside to 2. These values correspond to the effective indices inside and outside a silicon ridge waveguide on the silica substrate used in our experiment (to be described later). The outer radius of the spiral is set to $R_0$ = 106 µm, and the total length of the waveguide is $L$ = 4.8 mm. The input light excites all the waveguide modes with the same amplitude and random phase. We numerically solved light propagation in each mode of the spiral waveguide. The output field pattern is then obtained by summing the fields of all modes at the output port. By repeating the calculation at different wavelengths, we construct a transfer matrix that records the output intensity pattern as a function of the input wavelength.

In the spiral structure, the curving of the waveguide introduces mixing of guided modes. Also the sidewall roughness of the fabricated waveguide causes random mode mixing. Such mixing is usually detrimental to spectrometer operation as it suppresses the temporal spread of light [45] and reduces the spectral resolution [32]. In a multimode fiber with strong mode mixing, the spectral resolution scales as $\sqrt{L}$ instead of $L$, where $L$ is the length of the fiber. We simulated mode mixing in the spiral waveguide using a random matrix approach where the rate of mode mixing scales inversely with the local radius of curvature (see details in Supplement 1).

To separate the effects of mode mixing and evanescent coupling, we considered four cases that are shown in Fig. 1(b). The panel (i) is the transfer matrix in the absence of both mode mixing and evanescent coupling in a spiral waveguide. Each column represents the intensity pattern at the output port for light of wavelength $\lambda$, $I(x, \lambda)$, where $x$ is the spatial coordinate on the waveguide cross-section at the output port. We compute the spectral correlation function $C(\Delta\lambda) = \langle\langle I(x,\lambda)I(x,\lambda+\Delta\lambda)\rangle_\lambda / [\langle I(x,\lambda)\rangle_\lambda \langle I(x,\lambda+\Delta\lambda)\rangle_\lambda] - 1\rangle_x$, which is plotted by the dashed line in Fig. 1(c). The spectral correlation width $\delta\lambda$, i.e., the half-width-at-half-maximum (HWHM) of $C(\Delta\lambda)$, is 0.4 nm. This value provides an estimate of the spectral resolution, as a noticeable change in the intensity pattern is needed to distinguish between two wavelengths.

With evanescent coupling introduced to the spiral waveguide (no mode mixing), the output intensity pattern varies more rapidly with wavelength [panel (ii) of Fig. 1(c)]. The spectral correlation function is narrowed to a HWHM of 0.3 nm [dotted line in Fig. 1(c)], confirming that evanescent coupling improves the spectral resolution. Next we introduced mode mixing but neglect evanescent coupling. The output intensity pattern changes much more slowly with wavelength than in the previous two cases [panel (iii) in Fig. 1(b)]. The spectral correlation width $\delta\lambda$ increases to 0.9 nm [dash-dotted line in Fig. 1(c)], confirming that mode mixing decreases the spectral resolution. Finally, we include both evanescent coupling and mode mixing. The intensity pattern changes very quickly with wavelength [panel (iv) in Fig. 1(c)], in fact the change is faster than the case with evanescent coupling but no mode mixing. This observation is confirmed by the spectral correlation function shown by the solid line in Fig. 1(c), which features the narrowest width of 0.17 nm.

In contrast to its detrimental effect mentioned earlier, the mode mixing in the spiral waveguide improves the spectral resolution by enhancing the evanescent coupling for the lower-order modes. Although the evanescent coupling is weaker for the lower-order modes, the mode mixing can indirectly enhance the coupling by converting them to the higher-order modes (with stronger evanescent coupling) and then back. Therefore, the synergy between mode mixing and evanescent coupling leads to a significant enhancement of the spectral resolution.

We also investigated the scaling of the spectral resolution with the outer leads radius $R_0$ of the spiral. As detailed in the Supplement 1, when the number of spiral arms is fixed, the spectral correlation width $\delta\lambda$ scales as $1/R_0^2$ in the presence of mode mixing and evanescent coupling. This is in sharp contrast to the $1/R_0$ scaling when mode mixing and evanescent coupling are absent. Since the overall length $L$ of the spiral waveguide increases linearly with $R_0$, the combination of evanescent coupling and mode mixing results in a quadratic growth of the temporal spread and the spectral resolution with $L$.

## 3. SPECTROMETER OPERATION

Figure 2(a) is a conceptual schematic of a spiral spectrometer with integrated detectors on a silicon-on-insulator wafer. Light is coupled to the input waveguide from the edge of the chip via a lensed fiber. After propagating through the spiral, the output beam expands laterally in a tapered region before reaching the detector array. The lateral expansion increases the speckle size so that individual speckle grains can be resolved by the detectors.

Experimentally, we fabricated a series of multimode spiral waveguides via e-beam lithography and reactive ion etching [Fig. 2(b)]. The 220 nm thick silicon layer was fully etched to create an air gap between adjacent spiral arms. The air gap width was varied from 1 µm to 50 nm [Fig. 2(c, d)]. The outer radius of the spiral is $R_0$ = 250 µm, and the overall length of the waveguide is about 18 mm. In the initial prototype, we avoided the complication of integrating detectors on the silicon chip by scattering light out of the plane from the end facet of the tapered region. This allowed us to image the in-plane speckle pattern from the top using a 100× long working distance objective (numerical aperture NA = 0.65) and an InGaAs camera (Xenics Xeva-640, 640×480 pixels of 28×28 µm). The length of the tapered region was set to ∼60 µm to reduce the in-plane reflection from the groove to the spiral waveguide, while increasing the speckle size to match the resolution of our imaging system (∼ 1.5 µm).

In order to use the spiral waveguide as a spectrometer, we measured the output speckle pattern as a function of input wavelength [Fig. 3(a)]. This calibration is stored in a transfer matrix, $T$, relating the discretized spectral channels of input, S, to the spatial intensity distribution at the output, $I$, as $I = T \cdot S$ [30]. An example transfer matrix is presented in Fig. 3(b) for a spiral structure with 50 nm gap. The wavelength ranges from 1520 nm to 1522 nm with a step size of 2 pm.

From the measured transfer matrix, we calculated the spectral correlation function plotted in Fig. 3(c). The spiral waveguide with 1 µm gap experiences negligible evanescent coupling, and the spectral correlation width $\delta\lambda$ is 0.07 nm. As the gap narrows down to 100 nm, the evanescent coupling becomes significant, and $\delta\lambda$ decreases to 0.02 nm. A further reduction of the gap width to 50 nm further enhances the evanescent coupling, and $\delta\lambda$ drops to 0.01 nm.

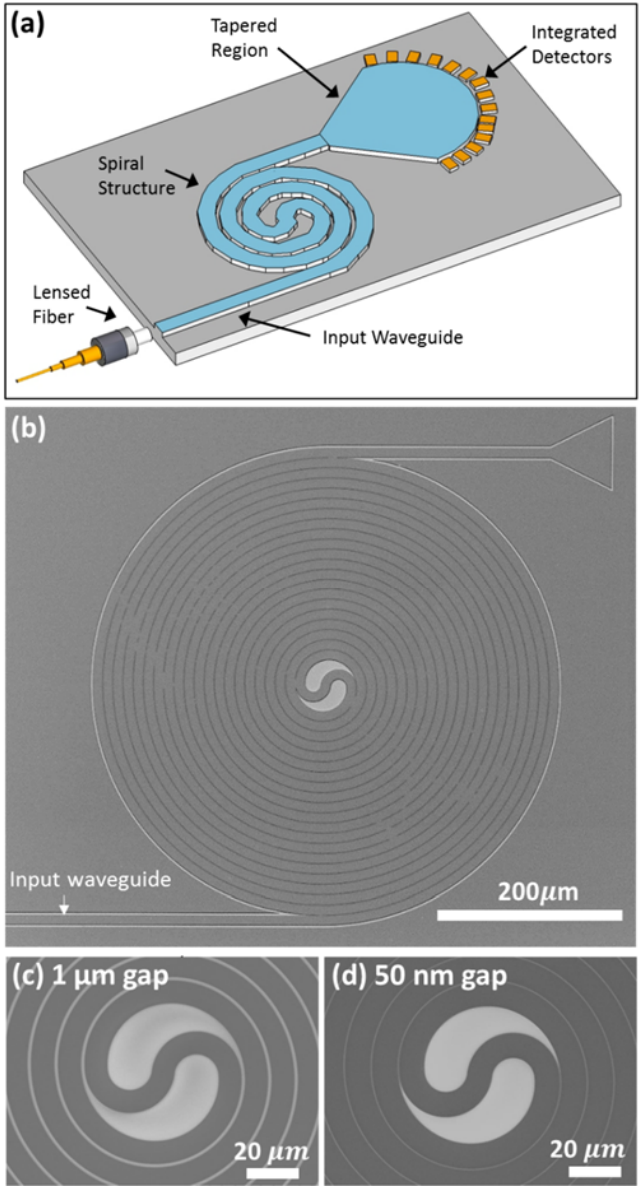

Fig. 2 Spiral spectrometer in a silicon-on-insulator wafer. (a) Schematic of a spiral spectrometer with integrated detectors. (a) Top-view SEM image of the silicon spiral waveguide that is 10 μm wide and 18 mm long. (b, c) Close-up SEM images showing the air gap between neighboring waveguide arms, and the gap width is 1 μm (b) or 50 nm (c).

While the spectral correlation width provides an estimate for the spectral resolution, the actual resolution is determined by the ability to distinguish two closely spaced spectral lines. We synthesized the output intensity pattern for an input spectrum by adding the speckle patterns measured sequentially at individual wavelengths, since optical signals at different wavelengths do not interfere. The spectrum $S$ was then reconstructed from the speckle pattern $I$ via nonlinear minimization of $\|I - T \cdot S\|^2 = \sum_j |I_j - \sum_i T_{j,i} S_i|^2$ [22,29,30]. The recovered spectrum in Fig. 3(d) shows that two lines separated by 0.01 nm in wavelength can be resolved by the spiral spectrometer with 50 nm wide gap.

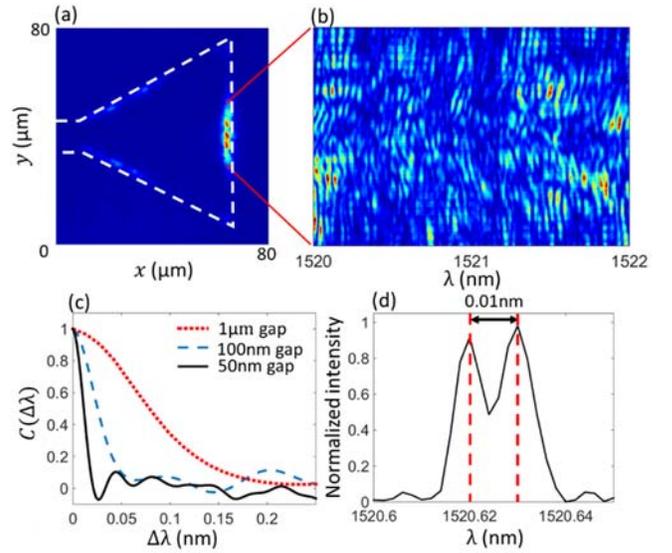

Fig. 3. Calibration and testing of the spiral spectrometer. (a) Top-view optical image of the tapered triangular region at the output port of the spiral waveguide (the boundary is marked by the white dashed line). (b) Measured spectral-spatial transfer matrix $T$ for a spiral waveguide with a 50 nm coupling gap, each column represents the intensity distribution on the end facet of the tapered region at a specific wavelength $I(x,\lambda)$. (c) The spectral correlation function $C(\Delta\lambda)$ obtained from the measured $I(x,\lambda)$ for three spirals with different gap width. The spectral correlation width $\delta\lambda$ [HWHM of $C(\Delta\lambda)$] is 0.07 nm for the 1μm gap, 0.02 nm for the 100 nm gap, and 0.01 nm for the 50 nm gap. (d) A reconstructed spectrum (black solid line) consisting of the two lines separated by 0.01 nm. Vertical red dotted lines mark the center wavelengths of the two lines.

Next we investigated the continuous bandwidth of the spiral spectrometer. Without prior information about the input spectrum, the number of independent spectral channels (producing uncorrelated speckle patterns) that can be reconstructed simultaneously is limited by the number of independent spatial channels $M$ in a speckle pattern $I(x,\lambda)$. To determine $M$, we applied the Karhunen-Loève decomposition [46,47] to the experimentally measured speckle pattern $I(x,\lambda)$. We built the spatial covariance matrix $C(x_1, x_2) = \langle I(x_1,\lambda)I(I(x_2,\lambda))\rangle_\lambda$ and computed its eigenvalues. As shown in Fig. 4(a), the kink in the semi-log plot of the eigenvalues gives the number of orthogonal spatial modes, $M$ = 40. The continuous bandwidth of the spiral spectrometer is $M \times \delta\lambda$, where the spectral width of each independent spectral channel is given by the spectral correlation width $\delta\lambda$. For the waveguide with a 1 μm gap, the bandwidth is 2.8 nm; for the 50 nm gap, the bandwidth decreases to 0.4 nm. This reduction reflects the trade-off between spectral resolution and bandwidth.

Figure 4(b) shows two continuous spectra over 0.4 nm bandwidth measured using the 50 nm gap spiral spectrometer. The two spectra have the same shape but different magnitude. After rescaling one of them, the two spectra overlap, which confirms the linearity of the spectral measurement. In Supplement 1, we show that the measured speckle intensity grows linearly with the input light intensity.

To enhance the operation bandwidth, we explored compressive sensing (CS) for spectrum recovery. The complex interference in the evanescently-coupled highly-multimode spiral waveguide produces diverse spectral features which are ideally suited for compressive sensing [48]. So far CS has not been applied to on-chip spectrometers, despite a recent proposal of using hundreds of different photonic nanostructures as random masks for CS spectroscopy [49]. In our scheme, only one structure is used together with multiple detectors to obtain random projections in a single shot measurement, which is more suitable for on-chip spectrometers.

First let us consider a sparse spectrum that consists of discrete lines. Since only a small number of spectral channels carry signals, the total number of independent spectral channels that can be reconstructed simultaneously may well exceed the number of spatial channels. To reconstruct the sparse spectrum, we minimized $\|I - T \cdot S\|^2 + \gamma |S|$, where $|S| = \sum_i |S_i|$ and $\gamma$ is a parameter. The additional term $\gamma|S|$ regularizes the sparsity of the solution [50]. The value of $\gamma$ is estimated from a cross-validation analysis [51], then fine-tuned by minimizing the difference between the reconstructed spectrum and the original spectrum [52].

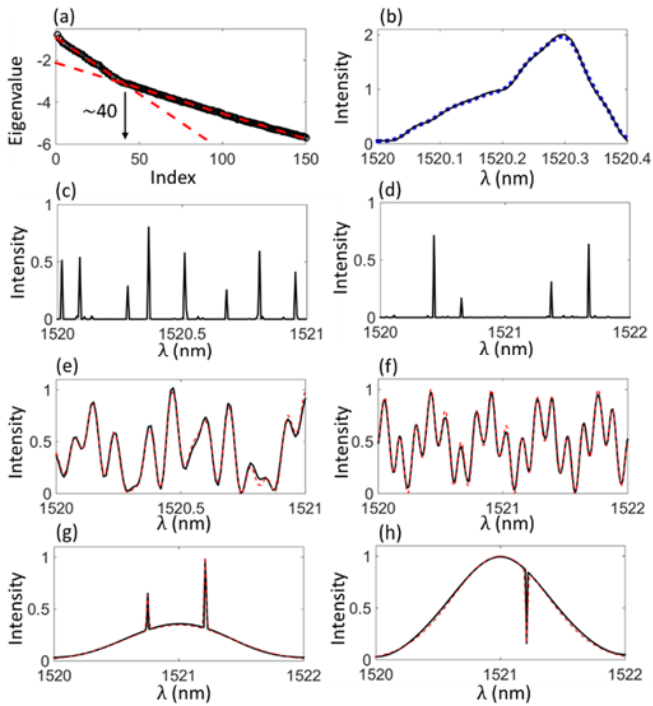

Fig. 4. Operation bandwidth of the spiral spectrometer. (a) Semi-log plot of the eigenvalues of the spatial covariance matrix $C(x_1, x_2)$ for the measured speckle pattern $I(x, \lambda)$, ordered by magnitude. The kink, indicated by the arrow, gives the number of orthogonal spatial modes, $M = 40$, in the output speckle pattern. (b) Two recovered spectra with identical shape but their magnitude differs by a factor of 2. After doubling the magnitude of the lower spectrum (blue dotted line), the two spectra overlap, verifying the linearity of measured spectra. (c, d) Reconstruction of a sparse spectrum that has 8 (c) or 4 (d) discrete lines of varying amplitude that are distributed over 166 (c) or 332 (d) spectral channels in a wavelength range of 1 nm (c) or 2 nm (d). (e, f) Reconstruction of a continuous spectrum with 6 (e) or 3 (f) DCT components over a 1 nm (e) or 2 nm (f) bandwidth. (g) Recovered spectrum having two sharp lines of different amplitude on top of a broad peak. (h) Recovered spectrum containing a narrow dip on a smooth broadband background.

Figure 4(c) shows a sparse spectrum that has 8 discrete lines distributed across 166 spectral channels, covering a wavelength range of 1 nm. A further increase in the sparsity enhances the bandwidth even more. As shown in Fig. 4(d), with only four spectral lines, 332 channels can be covered in a bandwidth of 2 nm. A more quantitative analysis of the number of lines that can be reconstructed and the effect of noise on the reconstruction accuracy are presented in the Supplement 1.

Next we apply the CS method to recover the "dense" spectrum that contains non-zero amplitude in every spectral channel. Such spectrum can be reconstructed by imposing the sparsity constraint in the appropriate basis. For example, a discrete cosine transform (DCT) of a smooth spectrum gives a limited number of low-frequency components. Therefore, we can reconstruct the continuous spectrum in the DCT domain, which becomes sparse [52]. Figure 4(e, f) shows two continuous spectra with different numbers of DCT components. A broader spectral range can be covered for fewer DCT components.

Furthermore, if the spectrum contains both sharp and smooth features, we can split the spectrum into two parts for reconstruction, one is sparse in the wavelength domain and another is sparse in the DCT domain [52]. In this way, we recover a spectrum that contains two sharp lines of different amplitude on top of a broad peak in Fig. 4(g). The bandwidth is determined by the number of discrete lines and DCT components in the spectrum. Figure 4(h) shows a spectrum that has a narrow dip on a smooth broadband background is recovered with high fidelity.

## 4. DISCUSSION AND CONCLUSION

In summary, we have designed and fabricated a high-resolution chip-scale spectrometer based on an evanescently coupled multimode spiral waveguide. Non-resonant, broadband enhancement of spectral resolution is achieved via evanescent coupling and mode mixing. Experimentally, we obtained a wavelength resolution of 0.01 nm at 1520 nm with a 250 μm radius spiral spectrometer based on a 10 μm wide silicon waveguide. Signals in 40 independent spectral channels are recovered simultaneously. We also showed that the operation bandwidth is significantly enhanced by using a compressive sensing algorithm to reconstruct sparse spectra.

A further increase of the bandwidth may be realized by increasing the waveguide width to accommodate more guided modes and thus allow simultaneous measurement of more independent spatial channels. However, for a given footprint, increasing the waveguide width will reduce the total length of the spiral waveguide and lower the spectral resolution, which represents a trade-off between bandwidth and resolution. In this initial demonstration, we aimed for high-resolution in a small footprint while compromising on bandwidth, but this trade-off may be adjusted for specific applications in the future. Despite its finite bandwidth, the spiral spectrometer can operate at any wavelength within the wide transparency window of silicon and silica, as long as the speckle patterns for these wavelengths are calibrated. In addition to providing high spectral resolution in a compact footprint, the spiral spectrometer is expected to have low loss. For example, silicon-based spiral waveguides have been realized with propagation loss of 3.6 dB/cm [53].

A final advantage of the spiral spectrometer is its robustness against fabrication imperfections. Although the spiral structure includes relatively small features (e.g., the air gap between the waveguides), modest variations from the design parameters would not significantly degrade the performance, as the spectrometer is calibrated after fabrication. Similar to most integrated photonic devices, the spiral spectrometer is sensitive to temperature change, which could alter the speckle pattern formed at a given wavelength. We performed numerical simulation to investigate the temperature sensitivity of the spiral spectrometer which is detailed in Supplement 1. The temperature sensitivity depends on the spectral resolution: the higher the resolution, the more sensitive the speckle pattern is to the temperature. For the spectrometer shown here with 0.01 nm resolution, a temperature change of 0.16 K causes a 50% decorrelation of the speckle pattern. Existing temperature controllers can provide sufficient thermal control and temperature stability for the spiral spectrometer to operate without recalibration. Alternately, multiple transfer matrices could be calibrated at varying temperatures, and the appropriate transfer matrix could then be selected to match the current chip temperature. Once photodetectors are integrated on-chip [54], the spiral spectrometer is expected to have a significant impact on low-cost portable sensing and to add functionality to lab-on-a-chip systems.

**Funding**. National Science Foundation (NSF) ECCS-1509361.

**Acknowledgment**. We thank Hemant Tagare for useful discussion on compressive sensing. Facilities used are supported by YINQE and NSF MRSEC Grant No. DMR-1119826.

See Supplement 1 for supporting content.

## REFERENCES

1. S. Janz, A. Balakrishnan, S. Charbonneau, P. Cheben, M. Cloutier, A. Delâge, K. Dossou, L. Erickson, M. Gao, P. A. Krug, B. Lamontagne, M. Packirisamy, M. Pearson, and D. Xu, "Planar Waveguide Echelle Gratings in Silica-On-Silicon," IEEE Photonics Technol. Lett. **16**, 503–505 (2004).
2. J. He, B. Lamontagne Delage, Andre, L. Erickson, M. Davies, and E. S. Koteles, "Monolithic Integrated Wavelength Demultiplexer Based on a Waveguide Rowland Circle Grating in InGaAsP / InP," J. Light. Technol. **16**, 631–638 (1998).
3. M. K. Smit and C. Van Dam, "PHASAR-based WDM-devices: Principles, design and applications," IEEE J. Sel. Top. Quantum Electron. **2**, 236–250 (1996).
4. P. Cheben, J. H. Schmid, a Delâge, a Densmore, S. Janz, B. Lamontagne, J. Lapointe, E. Post, P. Waldron, and D.-X. Xu, "A high-resolution silicon-on-insulator arrayed waveguide grating microspectrometer with sub-micrometer aperture waveguides.," Opt. Express **15**, 2299–2306 (2007).
5. Z. Shi and R. W. Boyd, "Fundamental limits to slow-light arrayed-waveguide-grating spectrometers," Opt. Express **21**, 7793–7798 (2013).
6. J. Zou, T. Lang, Z. Le, and J.-J. He, "Ultracompact silicon-on-insulator-based reflective arrayed waveguide gratings for spectroscopic applications," Appl. Opt. **55**, 3531–3536 (2016).
7. S. Babin, a. Bugrov, S. Cabrini, S. Dhuey, a. Goltsov, I. Ivonin, E.-B. Kley, C. Peroz, H. Schmidt, and V. Yankov, "Digital optical spectrometer-on-chip," Appl. Phys. Lett. **95**, 41105 (2009).
8. C. Peroz, C. Calo, a Goltsov, S. Dhuey, a Koshelev, P. Sasorov, I. Ivonin, S. Babin, S. Cabrini, and V. Yankov, "Multiband wavelength demultiplexer based on digital planar holography for on-chip spectroscopy applications.," Opt. Lett. **37**, 695–697 (2012).
9. G. Calafiore, A. Koshelev, S. Dhuey, A. Goltsov, P. Sasorov, S. Babin, V. Yankov, S. Cabrini, and C. Peroz, "Holographic planar lightwave circuit for on-chip spectroscopy," Light Sci Appl **3**, e203 (2014).
10. B. Momeni, E. S. Hosseini, and A. Adibi, "Planar photonic crystal microspectrometers in silicon-nitride for the visible range.," Opt. Express **17**, 17060–17069 (2009).
11. B. Gao, Z. Shi, and R. W. Boyd, "Design of flat-band superprism structures for on-chip spectroscopy," Opt. Express **23**, 6491 (2015).
12. B. E. Little, J. S. Foresi, G. Steinmeyer, E. R. Thoen, S. T. Chu, H. A. Haus, E. P. Ippen, L. C. Kimerling, and W. Greene, "Ultra-Compact Si – SiO Microring Resonator," IEEE Photonics Technol. Lett. **10**, 549–551 (1998).
13. A. Nitkowski, L. Chen, and M. Lipson, "Cavity-enhanced on-chip absorption spectroscopy using microring resonators.," Opt. Express **16**, 11930–11936 (2008).
14. B. B. C. Kyotoku, L. Chen, and M. Lipson, "Sub-nm resolution cavity enhanced micro- spectrometer," Opt. Express **18**, 102–107 (2010).
15. Z. Xia, A. A. Eftekhar, M. Soltani, B. Momeni, Q. Li, M. Chamanzar, S. Yegnanarayanan, and A. Adibi, "High resolution on-chip spectroscopy based on miniaturized microdonut resonators.," Opt. Express **19**, 12356–12364 (2011).
16. a Sharkawy, S. Shi, and D. W. Prather, "Multichannel wavelength division multiplexing with photonic crystals.," Appl. Opt. **40**, 2247–2252 (2001).
17. X. Gan, N. Pervez, I. Kymissis, F. Hatami, and D. Englund, "A high-resolution spectrometer based on a compact planar two dimensional photonic crystal cavity array," Appl. Phys. Lett. **100**, 231104 (2012).
18. A. C. Liapis, B. Gao, M. R. Siddiqui, Z. Shi, and R. W. Boyd, "On-chip spectroscopy with thermally tuned high-Q photonic crystal cavities," Appl. Phys. Lett. **108**, (2016).
19. Z. Xu, Z. Wang, M. Sullivan, D. Brady, S. Foulger, and A. Adibi, "Multimodal multiplex spectroscopy using photonic crystals.," Opt. Express **11**, 2126–2133 (2003).
20. T. W. Kohlgraf-Owens and A. Dogariu, "Transmission matrices of random media: means for spectral polarimetric measurements.," Opt. Lett. **35**, 2236–2238 (2010).
21. Q. Hang, B. Ung, I. Syed, N. Guo, and M. Skorobogatiy, "Photonic bandgap fiber bundle spectrometer.," Appl. Opt. **49**, 4791–4800 (2010).
22. B. Redding, S. F. Liew, R. Sarma, and H. Cao, "Compact spectrometer based on a disordered photonic chip," Nat. Photonics **7**, 746–751 (2013).
23. P. Wang and R. Menon, "Computational spectrometer based on a broadband diffractive optic.," Opt. Express **22**, 14575–14587 (2014).
24. M. Mazilu, T. Vettenburg, A. Di Falco, and K. Dholakia, "Random super-prism wavelength meter.," Opt. Lett. **39**, 96–99 (2014).
25. M. Chakrabarti, M. L. Jakobsen, and S. G. Hanson, "Speckle-based spectrometer," Opt. Lett. **40**, 3264–3267 (2015).
26. N. H. Wan, F. Meng, T. Schröder, R.-J. Shiue, E. H. Chen, and D. Englund, "High-resolution optical spectroscopy using multimode interference in a compact tapered fibre," Nat. Commun. **6**, 7762 (2015).
27. T. Yang, C. Xu, H. Ho, Y. Zhu, X. Hong, Q. Wang, Y. Chen, X. Li, X. Zhou, M. Yi, and W. Huang, "Miniature spectrometer based on diffraction in a dispersive hole array," Opt. Lett. **40**, 3217–3220 (2015).
28. J. Bao and M. G. Bawendi, "A colloidal quantum dot spectrometer," Nature **523**, 67–70 (2015).
29. B. Redding and H. Cao, "Using a multimode fiber as a high-resolution, low-loss spectrometer.," Opt. Lett. **37**, 3384–3386 (2012).
30. B. Redding, S. M. Popoff, and H. Cao, "All-fiber spectrometer based on speckle pattern reconstruction.," Opt. Express **21**, 6584–6600 (2013).
31. B. Redding, S. M. Popoff, Y. Bromberg, M. a Choma, and H. Cao, "Noise analysis of spectrometers based on speckle pattern

32. B. Redding, M. Alam, M. Seifert, and H. Cao, "High-resolution and broadband all-fiber spectrometers," Optica **1**, 175–180 (2014).
33. A. Densmore, D.-X. Xu, S. Janz, P. Waldron, T. Mischki, G. Lopinski, A. Delâge, P. Cheben, J. Lapointe, B. Lamontagne, and J. H. Schmid, "Spiral-path high-sensitivity silicon photonic wire molecular sensor with temperature-independent response.," Opt. Lett. **33**, 596–8 (2008).
34. D. X. Xu, a Densmore, a Delâge, P. Waldron, R. McKinnon, S. Janz, J. Lapointe, G. Lopinski, T. Mischki, E. Post, P. Cheben, and J. H. Schmid, "Folded cavity SOI microring sensors for high sensitivity and real time measurement of biomolecular binding.," Opt. Express **16**, 15137–15148 (2008).
35. D.-X. Xu, A. Delâge, R. McKinnon, M. Vachon, R. Ma, J. Lapointe, A. Densmore, P. Cheben, S. Janz, and J. H. Schmid, "Archimedean spiral cavity ring resonators in silicon as ultra-compact optical comb filters.," Opt. Express **18**, 1937–1945 (2010).
36. W. Guo and M. Digonnet, "Compact coupled resonators for slow-light sensor applications," Proc. SPIE **8636**, 863604 (2013).
37. W. Guo and M. J. F. Digonnet, "Coupled Spiral Interferometers," J. Light. Technol. **32**, 4162–4168 (2014).
38. W. Guo and M. J. F. Digonnet, "Coupled Spiral Interferometer Gyroscope," J. Light. Technol. **32**, 4360–4364 (2014).
39. T. Chen, H. Lee, J. Li, and K. J. Vahala, "A general design algorithm for low optical loss adiabatic connections in waveguides," Opt. Express **20**, 22819 (2012).
40. T. Chen, H. Lee, and K. J. Vahala, "Design and characterization of whispering-gallery spiral waveguides.," Opt. Express **22**, 5196–5208 (2014).
41. H. Lee, T. Chen, J. Li, O. Painter, and K. J. Vahala, "Ultra-low-loss optical delay line on a silicon chip.," Nat. Commun. **3**, 867 (2012).
42. D. Y. Oh, D. Sell, H. Lee, K. Y. Yang, S. a Diddams, and K. J. Vahala, "Supercontinuum generation in an on-chip silica waveguide.," Opt. Lett. **39**, 1046–1048 (2014).
43. H. Lee, M.-G. Suh, T. Chen, J. Li, S. a Diddams, and K. J. Vahala, "Spiral resonators for on-chip laser frequency stabilization.," Nat. Commun. **4**, 2468 (2013).
44. A. V Velasco, P. Cheben, P. J. Bock, A. Delâge, J. H. Schmid, J. Lapointe, S. Janz, M. L. Calvo, D.-X. Xu, M. Florjańczyk, and M. Vachon, "High-resolution Fourier-transform spectrometer chip with microphotonic silicon spiral waveguides.," Opt. Lett. **38**, 706–708 (2013).
45. D. J. Richardson, J. M. Fini, and L. E. Nelson, "Space-division multiplexing in optical fibres," Nat. Photonics **7**, 354 (2013).
46. H. Stark and J. W. Woods, eds., *Probability, Random Processes, and Estimation Theory for Engineers* (Prentice-Hall, Inc., 1986).
47. R. G. Ghanem and P. Spanos, *Stochastic Finite Elements: A Spectral Approach* (Springer-Verlag, 1991).
48. E. J. Candes and M. B. Wakin, "An Introduction To Compressive Sampling," IEEE Signal Process. Mag. **25**, 21–30 (2008).
49. Z. Wang and Z. Yu, "Spectral analysis based on compressive sensing in nanophotonic structures," Opt. Express **22**, 25608–25614 (2014).
50. S. Becker, J. Bobin, E. J. Candes, and E. Candès, "NESTA: a fast and accurate first-order method for sparse recovery," SIAM J. Imaging Sci. **4**, 1–39 (2011).
51. T. Hastie, R. Tibshirani, and J. Friedman, *The Elements of Statistical Learning* (Springer New York USA, 2001).
52. S. F. Liew, B. Redding, M. A. Choma, H. D. Tagare, and H. Cao, "Broadband multimode fiber spectrometer," Opt. Lett. **41**, 2029–2032 (2016).
53. Y. A. Vlasov and S. J. McNab, "Losses in single-mode silicon-on-insulator strip waveguides and bends," Opt. Express **12**, 1622–1631 (2004).
54. J. Michel, J. Liu, and L. C. Kimerling, "High-performance Ge-on-Si photodetectors," Nat Phot. **4**, 527–534 (2010).

reconstruction.," Appl. Opt. **53**, 410–417 (2014).

# Evanescently coupled multimode spiral spectrometer

BRANDON REDDING[†], SENG FATT LIEW[†], YARON BROMBERG, RAKTIM SARMA, HUI CAO [*]

*Department of Applied Physics, Yale University, New Haven, CT 06520*

*Corresponding author: hui.cao@yale.edu

†These authors contributed equally to this work.

This document provides supplementary information to "Evanescently coupled multimode spiral spectrometer". It includes a detailed description on the numerical simulation of the spiral spectrometer. The scaling of the spectral resolution and transmission with the spiral size is presented along with simulations of the temperature dependence of the device.   © 2014 Optical Society of America

http://  [supplementary document doi]

### Numerical modeling of spiral spectrometer

We performed numerical simulation of the interleaved Archimedean spirals. The local radius of curvature, $R$, is defined as a function of the azimuthal angle $\theta$ as $R(\theta) = R_0 - a\theta/\pi$, where $R_0$ denotes the radius of the spiral, $a$ is the center-to-center arm spacing (the sum of the waveguide width and the coupling gap), and $\theta$ increases from $0$ to $\theta_0$. The number of spiral arms is determined by $\theta_0$. The center position of the multimode waveguides are then defined in Cartesian coordinates as $x = \pm R(\theta)\cos(\theta)$, $y = \mp R(\theta)\sin(\theta)$, where the sign gives the two interleaved spirals. The two spirals are connected at the center by a S-bend that consists of two arcs of a circle [35, 39].

We modeled the wave propagation in the spiral waveguide using the approach in [36, 37]. First we ignored the mixing of waveguide modes with different order, which allowed us to simulate individual modes separately by solving the following two coupled equations:

$$\frac{dU_n(\theta)}{d\theta} = i\beta R(\theta) U_n(\theta) + \kappa_{n+1,n} U_{n+1}(\theta) + \kappa_{n-1,n} U_{n-1}(\theta)$$

$$\frac{dV_n(\theta)}{d\theta} = -i\beta R(\theta) V_n(\theta) - \kappa_{n+1,n} V_{n+1}(\theta) - \kappa_{n-1,n} V_{n-1}(\theta)$$

where $U_n$ and $V_n$ denote the forward (with respect to the input direction) and backward propagating waves in one waveguide mode in the $n^{th}$ arm of the spiral. $n$ ranges from $1$ (the outermost arm) to $N$ (the innermost arm), with $N$ given by $\theta_0$. $\beta$ is the propagation constant for this waveguide mode. $\kappa_{n+1,n}$ is the complex coefficient for evanescent coupling into the $n$th arm from the $(n+1)^{th}$ arm. Both propagation constant $\beta$ and coupling coefficient $\kappa$ are wavelength dependent. We consider only the coupling between the modes that have the same order and the same propagating direction in neighboring waveguides, as the coupling between modes with different order or opposite direction is negligible. For energy conservation and symmetry reasons, $\kappa_{n+1,n} = -(\kappa_{n,n+1})^* = i\kappa_n$, where $\kappa_n$ is a real number. Since the fields must be continuous at the junctures between adjacent arms, the continuity relations are $U_{j+1}(0) = U_j(2\pi) \exp[i\beta S_n]$, $V_{j+1}(0) = V_j(2\pi) \exp[-i\beta S_n]$, where $S_n$ is the length of the $n$th arm of the spiral.

The above equations were solved numerically by dividing the waveguide into many small segments. Within each segment, the waveguide may be approximated as a straight waveguide, because the local radius of curvature $R(\theta)$ is much larger than the length of the segment. The evanescent coupling length $L_c$ for two straight waveguides is obtained from the finite-difference frequency-domain calculation (COMSOL), and $\kappa_n = R_n/L_c$, where

$R_n$ is the radius of the $n$th arm of the spiral. The higher-order modes of the waveguide experience stronger evanescent coupling, thus having shorter $L_c$ and larger $\kappa_n$.

In the spiral structure, the curving of the waveguide introduces mode mixing. Also the sidewall roughness of the fabricated waveguide causes random mixing of the modes. We simulated mode mixing by introducing a random unitary matrix in between any two successive segments of the waveguide. The rate of mode mixing scales inversely with the local radius of curvature. All the modes are completely mixed after propagating through one arm of the spiral. In the center "S-bend", the modes are randomly mixed, as modeled by a random unitary matrix

In all four cases shown in Fig. 1(b,c) of the main text, the outer radius of the spiral is fixed to $R_0 = 106\,\mu m$, and the total length of the waveguide is $L$ = 4.8 mm. The evanescent coupling length $L_c$ varies from 0.55 mm for the lowest-order mode to 6 μm for the highest-order mode of the waveguide. Since the coupling lengths are shorter than the waveguide length, all the modes experience evanescent coupling while propagating through the spiral structure. However, the coupling is weaker for the lower-order modes; fortunately, the mode mixing can indirectly enhance the coupling rate by converting them to the higher-order modes (with stronger evanescent coupling) and then back. Therefore, the synergy between mode mixing and evanescent coupling leads to a significant enhancement of the spectral resolution.

We also investigated the scaling of the spectral resolution with the outer radius R0 of the spiral. Figure S1(a) plots the spectral correlation width δλ as a function of R0 when the number of spiral arms is fixed to N = 8. Without mode mixing and evanescent coupling, δλ scales as $1/R_0$. This behavior is expected, as the spiral length $L$ grows linearly with $R_0$. In the presence of mode mixing and evanescent coupling, the scaling changes to $1/R_0^2$, indicating the temporal spread of light grows as $R_0^2$ or $L^2$.

The evanescent coupling introduced to the interleaved spiral structure also causes a reduction in the total transmission, as part of the light propagates backward to the input port. To see how much the reduction is, we plot the total transmission $T$ in Fig. S1(b). When the coupling gap width is fixed, all waveguide modes experience more evanescent coupling as $R_o$ increases. Consequently, $T$ first drops and then levels off, eventually about half of the light is transmitted. To increase the transmission, the input port could be covered by a highly reflecting photonic crystal wall (having a full photonic bandgap in-plane) with a small opening (a defect waveguide for input light). Thus most of the light propagating back to the input port could be reflected to the spiral waveguide. This will enhance not only the transmission, but also the spectral resolution as the temporal spread of the transmitted light will further increase.

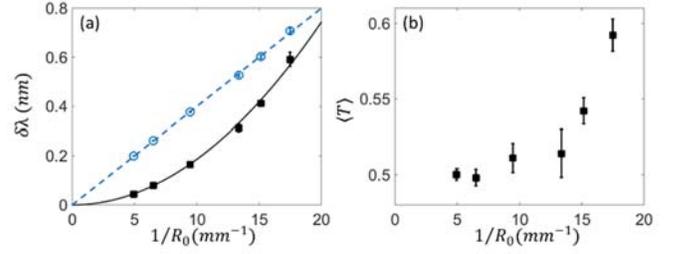

Fig. S1 Dependence of spectral resolution and total transmission on the size of spiral spectrometer. (a) Spectral correlation width δλ vs. the inverse of the outer radius, $1/R_0$, of the spiral waveguide with (solid black squares) and without (open blue circles) evanescent coupling and mode mixing. In the former case δλ scales as $1/R_0^2$ (black solid line), and in the latter δλ ~ $1/R_0$ (blue dashed line). (b) Total transmission vs. $1/R_0$ with evanescent coupling and mode mixing. The error bar is obtained from 50 ensembles of random mode mixing. Due to evanescent coupling, the transmission decreases with increasing $R_0$ and eventually saturates to 0.5.

## Linearity of Speckle Intensity

To check the linearity of the spectral measurements, we show in Fig. 4(b) of the main text two continuous spectra that have the same shape but differ by a factor of 2 in input intensity. After scaling the recovered spectrum from the weaker input by a factor 2, the two spectra overlap, verifying the linearity. We also measured the speckle pattern from the spiral waveguide at a single wavelength while increasing the input light intensity. Figure S2 shows the speckle intensity increases linearly with the input intensity.

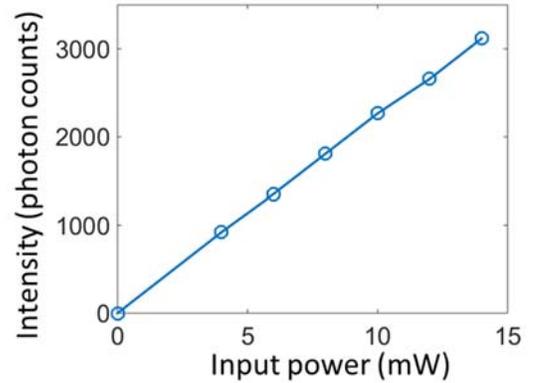

Fig. S2: Measured speckle intensity of a spiral spectrometer as a function of the input intensity at wavelength λ = 1520 nm. The spiral outer radius is 250 μm, the waveguide width is 10 μm, and the air gap between the neighboring spiral arm is 50 nm wide.

## Reconstruction of Sparse Spectrum

We used the compressive sensing algorithm to recover sparse spectra. The total number of independent spectral channels that can be reconstructed simultaneously is $P$, but only $H$ spectral channels carry signals and the other $P$-$H$ channels have zero amplitude. For such sparse spectra, $P$ can be significantly larger than the number of spatial channels $M$ when using compressive sensing, as shown in Fig. 4(c,d) of the main text.

To be more quantitative, we analyzed the number of spectral lines that can be reconstructed in a finite range of wavelength by fixing $P$ and $M$ while gradually increasing $H$. The reconstruction error $\varepsilon$ is determined by the difference in wavelength between the $H$ strongest lines in the recovered spectrum and those in the input spectrum. Figure S3 plots $\varepsilon$ averaged over 50 different spectra that have $H$ discrete lines of equal height and random spacing. We set $P = 332$, $M = 40$, and reconstructed the spectrum using the measured transfer matrix of the spiral spectrometer with 50 nm gap. The error $\varepsilon$ remains zero for $H$ up to 9, then started growing beyond that. This analysis gave the maximum number of discrete lines that could be reconstructed reliably.

The noise level in the measurement affects the number of spectral lines that could be accurately recovered. To simulate the measurement noise, we added white noise to an initially "noise-less" speckle pattern that was synthesized from the same transfer matrix used by the reconstruction algorithm. The SNR is defined as the ratio of the integrated intensity of the signal (without noise) to that of the noise. The lines in Fig. S3 represent the reconstruction error $\varepsilon$ for three different levels of SNR. A higher SNR allows more spectral lines to be recovered accurately. The SNR in our measurement is about 20 dB, and the reconstruction error matches the simulation result for SNR = 20 dB.

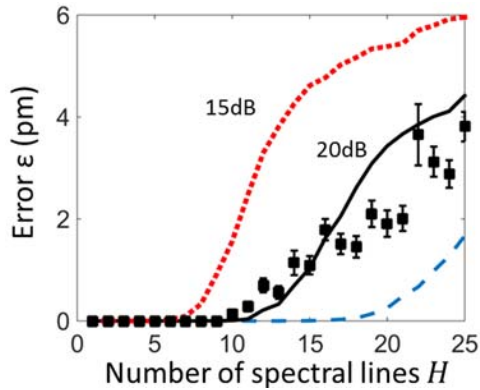

Fig. S3: The reconstruction error $\varepsilon$ as a function of the number of discrete lines $H$ present within the 2nm bandwidth ($P = 332$, $M = 40$). The solid squares are experimental data averaged over 50 different spectra. The lines are simulated results for three different SNR levels. More discrete lines can be recovered with higher SNR.

### Temperature Sensitivity

To quantitatively estimate the temperature sensitivity of the spiral spectrometer, we numerically simulated the change of output speckle pattern with temperature. We assume that the temperature change only modifies the real part of the refractive index of the silicon. Therefore, in the simulation, we vary the refractive index of the silicon gradually and calculate the speckle pattern at a fixed wavelength of 1.5 μm. Then we compute the correlation of the speckle pattern to the original one, and estimate the amount of refractive index change $\Delta n$ that causes 50% decorrelation of the speckle pattern. From the value of $\Delta n$, we find the corresponding change of temperature $\Delta T$ using the thermo-optic coefficient of Si at T = 295 K and λ = 1.5μm, where dn/dT = 1.87E-04 [1]. The $\Delta T$ can be taken as an approximate measure of the temperature sensitivity.

We simulated the temperature sensitivity for 3 spiral spectrometers with different waveguide length and spectral resolution. As shown in Table S1, the spiral with finer spectral resolution is more sensitive to the temperature change. The temperature sensitivity scales linearly with the spectral resolution. For the spiral spectrometer with 0.012 nm resolution, which is close to our experimentally measured resolution in Fig. 3(c), a temperature change of 0.16 K is sufficient to decorrelate the speckle pattern by 50%. Existing temperature controllers can provide excellent temperature stability (e.g. Thorlabs TED4015 provides stability of <0.002 K) and would thus provide sufficient thermal control for the spiral spectrometer to operate without recalibration. Alternatively, the thermal stability requirement could be loosened by calibrating the speckle patterns at multiple temperatures and selecting the transfer matrix for the current chip temperature. Typically the temperature variation is slower than the spectrum measurement, and the complexity of reconstruction algorithm would not increase, provided the chip temperature is constantly monitored.

| Total length of spiral $L$ (mm) | Spectral correlation function width $\Delta\lambda$ (nm) | Refractive index correlation function width $\Delta n$ | Corresponding change in temperature $\Delta T$ (K) |
|---|---|---|---|
| 4.83 | 0.164 | $39 \times 10^{-5}$ | 2.10 |
| 9.7 | 0.044 | $11 \times 10^{-5}$ | 0.59 |
| 19.9 | 0.012 | $2.9 \times 10^{-5}$ | 0.16 |

Table S1. Temperature sensitivity of the spiral spectrometer. The amount of refractive index change $\Delta n$ and the corresponding temperature change $\Delta T$ to cause 50% decorrelation of the speckle pattern are estimated for three spiral spectrometers of different waveguide length and spectral resolution.

### Comparison of On-chip Spectrometers

Since our spiral spectrometer is designed to be a completely on-chip, integrated spectrometer, we compare its performance to other spectrometers that can operate fully on-chip and would thus compete for the same application space.

| Spectrometer | Footprint | Resolution | Bandwidth | reference |
|---|---|---|---|---|
| Echelle grating | 18mm x 17mm | 0.8 nm | 40 nm (1530 -1568nm) | S. Janz, et al., IEEE Photon. Tech. Lett. 16, 503 (2004) |
| | 9mm x 6mm | 0.5 nm | 15nm (at 845nm) | X. Ma, et al. IEEE Photon. Journal 5,6600807 (2013) |
| Arrayed waveguide grating | 8mm x 8 mm | 0.2 nm | 10 nm (1539-1549 nm) | P. Cheben, et al., Opt. Express. 15, 2299 (2007) |
| | 860um x 240um | 1.5nm | 40nm (at 1500nm) | J. Zou, et al.," Appl. Opt. 55, 3531 (2016) |
| Digital planar holography | ~2mm² | 0.15 nm | 148 nm (at 600nm) | G. Calafiore, et al., Light: Science & Applications 3, e203 (2014). |
| Photonic crystal | 80um x 220um | 8 nm (single lines at ~10 pm) | 50 nm (1560-1610 nm) | B. Momeni, et al., Opt. Commun. 282, 3168 (2009) |
| Microdonut | 1 mm² | 0.6 nm | 50 nm (1550-1600 nm) | Z. Xia, et al., Opt. Express 19 12356 (2011) |
| Random Spectrometer | 100um x 50um | 0.75 nm | 25 nm (1500-1525nm) | Redding et al., Nature Photonics 7, 746 (2013) |
| Fourier-transform spiral waveguide | 2mm x 2mm | 0.042 nm | 0.75 nm (at 1500nm) | A. V. Velasco, et al., Opt. Lett. 38, 706 (2013) |

Table S2. Comparison of the footprint, wavelength resolution and bandwidth of our spiral spectrometer to the existing on-chip spectrometers.

**Reference:**

[1] B. J. Frey, D. B. Leviton, and T. J. Madison. *SPIE Astronomical Telescopes+ Instrumentation.* International Society for Optics and Photonics, pp. 62732J, 2006.